\documentclass{article}

\usepackage{latexsym}
\usepackage{graphicx}

\def\gsize{0.35} 
\def\gs2{0.60} 
\def\g3{0.9} 

\title{Statistical Mechanics of Online Learning for Ensemble Teachers}

\author{Seiji MIYOSHI\thanks{
Department of Electronic Engineering, 
Kobe City College of Technology, 
8--3 Gakuen-higashimachi, Nishi-ku, Kobe-shi, 651--2194 
E-mail address: miyoshi@kobe-kosen.ac.jp}
and Masato OKADA\thanks{
Division of Transdisciplinary Sciences, 
Graduate School of Frontier Sciences, 
The University of Tokyo, 
5--1--5 Kashiwanoha, Kashiwa-shi, Chiba, 277--8561, 
RIKEN Brain Science Institute, 
2--1 Hirosawa, Wako-shi, Saitama, 351--0198 
JST PRESTO,
5--1--5 Kashiwanoha, Kashiwa-shi, Chiba, 277--8561
}}

\begin{document}


\maketitle

\section*{Abstract}
We analyze the generalization performance of a student 
in a model composed of linear perceptrons: 
a true teacher, ensemble teachers, and the student. 
Calculating the generalization error 
of the student analytically using statistical mechanics
in the framework of on-line learning, 
it is proven that
when learning rate $\eta <1$,
the larger the number $K$ and the variety of 
the ensemble teachers are,
the smaller the generalization error is.
On the other hand, when $\eta >1$,
the properties are completely reversed.
If the variety of the ensemble teachers is rich enough, 
the direction cosine between the true teacher and the 
student becomes unity in the limit of $\eta \rightarrow 0$ and
$K \rightarrow \infty$.

keywords: ensemble teachers, on-line learning, generalization error,
statistical mechanics, learning rate

\section{Introduction}
Learning is to infer the underlying rules that dominate 
data generation using observed data.
Observed data are input-output pairs from a teacher
and are called examples.
Learning can be roughly classified into batch learning and
on-line learning \cite{Saad}.
In batch learning, given examples are used repeatedly.
In this paradigm, a student becomes to give correct answers
after training if the student has adequate freedom.
However, it is necessary to have a long amount of time and 
a large memory in which to store many examples.
On the contrary, in online learning examples used once are discarded.
In this case, a student cannot give correct answers 
for all examples used in training.
However, there are merits, for example,
a large memory for storing many examples isn't necessary,
and it is possible to follow a time variant teacher. 

Recently, we \cite{Hara,PRE} 
analyzed the generalization performance
of ensemble learning
\cite{Abe,Krogh,Urbanczik}
in a framework of on-line learning
using a statistical mechanical method \cite{Saad,NishimoriE}.
Using the same method, 
we also analyzed the generalization performance
of a student supervised by a moving teacher that goes around
a true teacher\cite{JPSJ2006}.
As a result, 
it was proven that
the generalization error of a student
can be smaller than
a moving teacher,
even if the student only uses examples 
from the moving teacher.
In an actual human society, a teacher
observed by a student
doesn't always present the correct answer.
In many cases, the teacher is learning
and continues to change.
Therefore, the analysis of such a model
is interesting for considering the 
analogies between statistical learning theories 
and an actual human society.

On the other hand, 
in most cases in an actual human society 
a student can observe examples from two or more teachers
who differ from each other.
Therefore, we analyze the generalization performance
of such a model and discuss the 
use of imperfect teachers in this paper.
That is, we consider a true teacher and
$K$ teachers called ensemble teachers who 
exist around the true teacher.
A student uses input-output pairs from ensemble teachers
in turn or randomly.
In this paper, we treat a model in which
all of the true teacher, the ensemble teachers and the student
are linear perceptrons\cite{Hara} with noises.
We obtain order parameters and generalization errors
analytically 
in the framework of on-line learning 
using a statistical mechanical method.
As a result, 
it is proven that
when student's learning rate $\eta <1$,
the larger the number $K$ and the variety of 
the ensemble teachers are,
the smaller the student's generalization error is.
On the other hand, when $\eta >1$,
the properties are completely reversed.
If the variety of ensemble teachers is rich enough, 
the direction cosine between the true teacher and the 
student becomes unity in the limit of $\eta \rightarrow 0$ and
$K \rightarrow \infty$.

\section{Model}
In this paper, we consider a true teacher, $K$ ensemble teachers
and a student.
They are all linear perceptrons with connection weights
$\mbox{\boldmath $A$}$, 
$\mbox{\boldmath $B$}_k$ and
$\mbox{\boldmath $J$}$, respectively.
Here, $k=1,\ldots,K$.
For simplicity, the connection weight of the true teacher,
the ensemble teachers and the student
are simply called the true teacher, the ensemble teachers and
the student, respectively.
True teacher $\mbox{\boldmath $A$}=\left(A_1,\ldots,A_N\right)$,
ensemble teachers 
$\mbox{\boldmath $B$}_k=\left(B_{k1},\ldots,B_{kN}\right)$,
student
$\mbox{\boldmath $J$}=\left(J_1,\ldots,J_N\right)$
and input 
$\mbox{\boldmath $x$}=\left(x_1,\ldots,x_N\right)$
are $N$ dimensional vectors.
Each component $A_i$ of $\mbox{\boldmath $A$}$
is drawn from ${\cal N}(0,1)$ independently and fixed,
where ${\cal N}(0,1)$ denotes Gaussian distribution with
a mean of zero and variance unity.
Some components $B_{ki}$
are equal to $A_i$ multiplied by --1,
the others are equal to $A_i$.
Which component $B_{ki}$ is equal to $-A_i$
is independent from the value of $A_i$.
Hence, $B_{ki}$ also obeys ${\cal N}(0,1)$.
$B_{ki}$ is also fixed.
The direction cosine between 
$\mbox{\boldmath $B$}_k$ and
$\mbox{\boldmath $A$}$ is $R_{Bk}$
and that between
$\mbox{\boldmath $B$}_k$ and
$\mbox{\boldmath $B$}_{k'}$
is $q_{kk'}$.
Each of the components $J_i^0$ 
of the initial value $\mbox{\boldmath $J$}^0$
of $\mbox{\boldmath $J$}$
are drawn from ${\cal N}(0,1)$ independently.
The direction cosine between 
$\mbox{\boldmath $J$}$ and 
$\mbox{\boldmath $A$}$ is  $R_{J}$
and that between
$\mbox{\boldmath $J$}$ and
$\mbox{\boldmath $B$}_{k}$ is $R_{BkJ}$.
Each component $x_i$ of $\mbox{\boldmath $x$}$
is drawn from ${\cal N}(0,1/N)$ independently.
Thus,
\begin{eqnarray}
\left\langle A_i\right\rangle &=& 0, \ \ 
\left\langle \left(A_i\right)^2\right\rangle=1, \\
\left\langle B_{ki}\right\rangle &=& 0, \ \ 
\left\langle \left(B_{ki}\right)^2\right\rangle=1,\\
\left\langle J_i^0\right\rangle &=& 0, \ \ 
\left\langle \left(J_i^0\right)^2\right\rangle=1,\\
\left\langle x_i\right\rangle &=& 0, \ \ 
\left\langle \left(x_i\right)^2\right\rangle=\frac{1}{N}, \\
R_{Bk}&=&\frac{\mbox{\boldmath $A$}\cdot\mbox{\boldmath $B$}_k}{\|\mbox{\boldmath $A$}\|\|\mbox{\boldmath $B$}_k\|}, \ \ 
q_{kk'}=\frac{\mbox{\boldmath $B$}_k \cdot \mbox{\boldmath $B$}_{k'}}{\|\mbox{\boldmath $B$}_k \|\| \mbox{\boldmath $B$}_{k'}\|}, \\
R_{J}&=&\frac{\mbox{\boldmath $A$}\cdot\mbox{\boldmath $J$}}{\|\mbox{\boldmath $A$}\|\|\mbox{\boldmath $J$}\|}, \ \ 
R_{BkJ}=\frac{\mbox{\boldmath $B$}_k \cdot \mbox{\boldmath $J$}}{\|\mbox{\boldmath $B$}_k \|\| \mbox{\boldmath $J$}\|},
\end{eqnarray}
where $\langle \cdot \rangle$ denotes a mean.

\begin{figure}[htbp]
\vspace{3mm}
\begin{center}
\includegraphics[width=\gsize\linewidth,keepaspectratio]{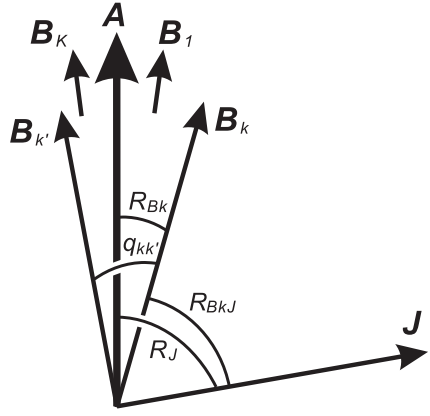}
\caption{True teacher $\mbox{\boldmath $A$}$,
ensemble teachers $\mbox{\boldmath $B$}_k$
and student $\mbox{\boldmath $J$}$.
$q_{kk'}, R_J, R_{Bk}$ and $R_{BkJ}$ are direction cosines.}
\label{fig:ABJ}
\end{center}
\end{figure}

Figure \ref{fig:ABJ} illustrates
the relationship among true teacher
$\mbox{\boldmath $A$}$,
ensemble teachers $\mbox{\boldmath $B$}_k$,
student $\mbox{\boldmath $J$}$
and direction cosines
$q_{kk'}, R_{Bk}, R_J$ and $R_{BkJ}$.

In this paper, the thermodynamic limit $N\rightarrow \infty$
is also treated. Therefore,
\begin{equation}
\|\mbox{\boldmath $A$}\|=\sqrt{N},\ \ 
\|\mbox{\boldmath $B$}_k\|=\sqrt{N},\ \ 
\|\mbox{\boldmath $J$}^0\|=\sqrt{N},\ \ 
\|\mbox{\boldmath $x$}\|=1.
\label{eqn:xBJ}
\end{equation}
Generally, norm $\|\mbox{\boldmath $J$}\|$
of the student
changes as time step proceeds.
Therefore, ratios $l^m$ of the norm to $\sqrt{N}$
are introduced and called the length of 
the student. That is,
$\|\mbox{\boldmath $J$}^m\|=l^m\sqrt{N}$,
where $m$ denotes the time step.

The outputs of the true teacher, the ensemble teachers, 
and the student are
$y^m+n_A^m$, $v_k^m+n_{Bk}^m$ and
$u^ml^m+n_J^m$, respectively.
Here,
\begin{eqnarray}
y^m &=&
 \mbox{\boldmath $A$}\cdot \mbox{\boldmath $x$}^m, \label{eqn:y}\\
v_k^m &=& 
 \mbox{\boldmath $B$}_k\cdot \mbox{\boldmath $x$}^m, \label{eqn:v}\\
u^m l^m &=& 
 \mbox{\boldmath $J$}^m\cdot \mbox{\boldmath $x$}^m, \label{eqn:u}\\
n_{A}^m &\sim& {\cal N}\left(0,\sigma_{A}^2\right),\\
n_{Bk}^m &\sim& {\cal N}\left(0,\sigma_{Bk}^2\right),\\
n_J^m &\sim& {\cal N}\left(0,\sigma_J^2\right).
\end{eqnarray}
That is,
the outputs of the true teacher, 
the ensemble teachers and the student
include independent Gaussian noises with variances of 
$\sigma_{A}^2, \sigma_{Bk}^2$, and $\sigma_J^2$, respectively.
Then, $y^m$, $v^m$, and $u^m$
of Eqs. (\ref{eqn:y})--(\ref{eqn:u})
obey Gaussian distributions with a mean of zero and 
variance unity.

Let us define error $\epsilon_{Bk}$ between 
true teacher $\mbox{\boldmath $A$}$
and each member $\mbox{\boldmath $B$}_k$ of 
the ensemble teachers
by the squared errors of their outputs:
\begin{equation}
\epsilon_{Bk}^m \equiv 
\frac{1}{2}\left( y^m+n_{A}^m - v_k^m -n_{Bk}^m\right)^2.
\label{eqn:hate}
\end{equation}

In the same manner,
let us define error $\epsilon_{BkJ}$ between 
each member $\mbox{\boldmath $B$}_k$ of 
the ensemble teachers
and student $\mbox{\boldmath $J$}$
by the squared errors of their outputs:
\begin{equation}
\epsilon_{BkJ}^m \equiv 
\frac{1}{2}\left( v_k^m+n_{Bk}^m-u^ml^m-n_J^m\right)^2.
\label{eqn:e}
\end{equation}

Student $\mbox{\boldmath $J$}$ 
adopts the gradient method as a learning rule
and uses
input $\mbox{\boldmath $x$}$
and an output
of one of the $K$ ensemble teachers
$\mbox{\boldmath $B$}_k$
in turn or randomly for updates.
That is,
\begin{eqnarray}
\mbox{\boldmath $J$}^{m+1}
&=& \mbox{\boldmath $J$}^{m} 
   -\eta \frac{\partial \epsilon_{BkJ}^m}{\partial \mbox{\boldmath $J$}^{m}}\\
&=&  \mbox{\boldmath $J$}^{m} 
   +\eta \left( v_k^m+n_{Bk}^m-u^ml^m-n_J^m\right)
   \mbox{\boldmath $x$}^{m},\label{eqn:Jupdate}
\end{eqnarray}
where $\eta$ denotes the learning rate of the student
and is a constant number.
In cases where the student uses 
$K$ ensemble teachers in turn,
$k=\mbox{mod}\left(m,K\right)+1$.
Here, $\mbox{mod}\left(m,K\right)$ denotes
the remainder of $m$ divided by $K$.
On the other hand, in random cases,
$k$ is a uniform random integer that 
takes one of $1,2,\ldots,K$.

Generalizing the learning rules, 
Eq. (\ref{eqn:Jupdate})
can be expressed as
\begin{eqnarray}
\mbox{\boldmath $J$}^{m+1}
&=& \mbox{\boldmath $J$}^{m}
 +f_k
 \mbox{\boldmath $x$}^{m}\\
&=& \mbox{\boldmath $J$}^{m}
 +f\left( v_k^m+n_{Bk}^m,u^ml^m+n_J^m\right)
 \mbox{\boldmath $x$}^{m}, \label{eqn:J}
\end{eqnarray}
where $f$ denotes a function
that represents the update amount and 
is determined by the learning rule.

In addition, let us define error $\epsilon_J$ between 
true teacher $\mbox{\boldmath $A$}$ and 
student $\mbox{\boldmath $J$}$ by 
the squared error of their outputs:
\begin{equation}
\epsilon_J^m \equiv 
\frac{1}{2}\left( y^m+n_{A}^m -u^ml^m-n_J^m\right)^2.
\label{eqn:bare}
\end{equation}

\section{Theory}
\subsection{Generalization error}
One purpose of a statistical learning theory
is to theoretically obtain generalization errors.
Since generalization error is the mean of errors 
for the true teacher 
over the distribution of new input and noises,
generalization error $\epsilon_{Bkg}$ of 
each member $\mbox{\boldmath $B$}_k$ of the ensemble teachers
and $\epsilon_{Jg}$ of student
$\mbox{\boldmath $J$}$
are calculated as follows.
Superscripts $m$, which represent the time steps, 
are omitted for simplicity unless stated otherwise.
\begin{eqnarray}
\epsilon_{Bkg}
&=& \int d\mbox{\boldmath $x$} dn_{A} dn_{Bk} 
         P\left(\mbox{\boldmath $x$}, n_{A}, n_{Bk}\right)
         \epsilon_{Bk}
         \\
&=& \int dy dv_k  dn_{A} dn_{Bk} 
    P\left(y, v_k, n_{A}, n_{Bk}\right)
    \frac{1}{2}
    \left( y+n_{A} - v_k -n_{Bk}\right)^2 \\
&=& \frac{1}{2}
    \left(
    -2R_{Bk} + 2 + \sigma_{A}^2 + \sigma_{Bk}^2
    \right),
    \label{eqn:hateg} \\
\epsilon_{Jg} 
&=& \int d\mbox{\boldmath $x$} dn_{A} dn_J 
         P\left(\mbox{\boldmath $x$}, n_{A}, n_J\right)
         \epsilon_J
         \\
&=& \int dy du dn_{A} dn_J 
    P\left(y, u, n_{A}, n_J\right)
    \frac{1}{2}
    \left(y+n_{A} - ul - n_J \right)^2 \\
&=& \frac{1}{2}
    \left(
    -2R_J l + l^2 + 1 + \sigma_{A}^2 + \sigma_J^2
    \right).
    \label{eqn:eJg}
\end{eqnarray}

Here, integrations have been executed 
using the following:
$y$, $v_k$ and $u$ obey ${\cal N}(0,1)$.
The covariance between $y$ and $v_k$ is $R_{Bk}$,
that between $v_k$ and $u$ is $R_{BkJ}$,
and that between $y$ and $u$ is $R_J$.
All $n_{A}$, $n_{Bk}$, and $n_J$
are independent from other probabilistic variables.

\subsection{Differential equations for order parameters and
their analytical solutions}
To simplify analysis, the following auxiliary order parameters
are introduced:
\begin{eqnarray}
r_J      &\equiv& R_J     l, \label{eqn:defrJ} \\
r_{BkJ}  &\equiv& R_{BkJ} l. \label{eqn:defrBkJ}
\end{eqnarray}

Simultaneous differential equations
in deterministic forms \cite{NishimoriE}, 
which describe the dynamical behaviors of order parameters,
have been obtained based on self-averaging
in the thermodynamic limits as follows:
\begin{eqnarray}
\frac{dr_{BkJ}}{dt}&=& 
\frac{1}{K}\sum_{k'=1}^K \langle f_{k'} v_k\rangle, \label{eqn:drBkJdt}\\
\frac{dr_J}{dt}&=&
\frac{1}{K}\sum_{k=1}^K \langle f_k y\rangle, \label{eqn:drJdt}\\
\frac{dl}{dt}&=&
\frac{1}{K}\sum_{k=1}^K
\left(\langle f_k u\rangle+\frac{1}{2l}\langle f_k^2 \rangle \right).
\label{eqn:dldt}
\end{eqnarray}
Here, dimension $N$ has been treated 
to be sufficiently greater than
the number of ensemble teachers $K$.
Time $t=m/N$, that is, 
time step $m$ normalized by dimension $N$.
Note that the above differential equations
are identical whether 
the $K$ ensemble teachers
are used in turn or randomly.

Since linear perceptrons are treated in this paper,
the sample averages that appeared in the above 
equations can be easily calculated as follows:
\begin{eqnarray}
\langle f_k u \rangle &=& \eta \left(\frac{r_{BkJ}}{l}-l\right), 
\label{eqn:fku}\\
\langle f_k^2 \rangle &=& 
\eta^2 \left(l^2-2r_{BkJ}+1+\sigma_{Bk}^2+\sigma_J^2\right), \\
\langle f_k y \rangle &=& \eta \left(R_{Bk} -r_J\right), \\
\frac{1}{K}\sum_{k'=1}^K \langle f_{k'} v_k \rangle
&=& \eta \left(-r_{BkJ}+\frac{1}{K}\sum_{k'=1}^Kq_{kk'}\right).
\end{eqnarray}

Since all components $A_i$, $J_i^0$ 
of true teacher $\mbox{\boldmath $A$}$,
and the initial student $\mbox{\boldmath $J$}^0$
are drawn from ${\cal N}(0,1)$ independently
and because the thermodynamic limit $N\rightarrow \infty$
is also treated,
they are orthogonal to each other in the initial state.
That is,
\begin{equation}
R_J=0\ \mbox{when}\ t=0.
\label{eqn:Rinit}
\end{equation}

In addition,
\begin{equation}
l=1\ \mbox{when}\ t=0.
\label{eqn:linit}
\end{equation}

By using Eqs. (\ref{eqn:fku})--(\ref{eqn:linit}),
simultaneous differential equations 
Eqs. (\ref{eqn:drBkJdt})--(\ref{eqn:dldt})
can be solved analytically as follows:
\begin{eqnarray}
r_{BkJ}&=&\frac{1}{K}\sum_{k'=1}^K q_{kk'} \left(1-e^{-\eta t}\right), 
\label{eqn:rBkJ} \\
r_J &=&\frac{1}{K}\sum_{k=1}^K R_{Bk} \left(1-e^{-\eta t}\right), 
\label{eqn:rJ} \\
l^2 &=& 
\frac{1}{2-\eta}
\left[2\left(1-\eta\right)\bar{q}+\eta \left(1+\bar{\sigma}_B^2+
\sigma_J^2\right)\right]\nonumber \\
&+& \left[1+\frac{1}{2-\eta}
\left(\eta \left(1+\bar{\sigma}_B^2+\sigma_J^2\right)
-2\bar{q}\right)\right]e^{\eta\left(\eta-2\right)t} - 2\bar{q}e^{-\eta t},\label{eqn:l2} 
\end{eqnarray}
where
\begin{eqnarray}
\bar{q} &=& \frac{1}{K^2}\sum_{k=1}^K \sum_{k'=1}^K q_{kk'}, 
\label{eqn:barq} \\
\bar{\sigma}_B^2 &=& \frac{1}{K}\sum_{k=1}^K \sigma_{Bk}^2.
\label{eqn:barsigma2B} 
\end{eqnarray}

\section{Results and Discussion}
In this section, 
we treat the case where direction cosines $R_{Bk}$ between 
the ensemble teachers and the true teacher,
direction cosines $q_{kk'}$ among the ensemble teachers
and variances $\sigma_{Bk}^2$ of the noises of 
ensemble teachers are uniform.
That is,
\begin{eqnarray}
R_{Bk}         &=& R_B,\ \  k=1,\ldots,K,\\
q_{kk'}
&=&\left\{
\begin{array}{ll}
q,           & k\neq k' , \\
1,           & k=k',
\end{array}
\right.\label{eqn:qkk} \\
\sigma_{Bk}^2  &=& \sigma_B^2.
\label{eqn:}
\end{eqnarray}

In this case,
Eqs. (\ref{eqn:barq}) and (\ref{eqn:barsigma2B})
are expressed as
\begin{eqnarray}
\bar{q}&=&q+\frac{1-q}{K},\label{eqn:barq2} \\
\bar{\sigma}_B^2&=&\sigma_B^2.\label{eqn:barsigma2B2} 
\end{eqnarray}

\begin{figure}[htbp]
\begin{center}
\includegraphics[width=\gs2\linewidth,keepaspectratio]{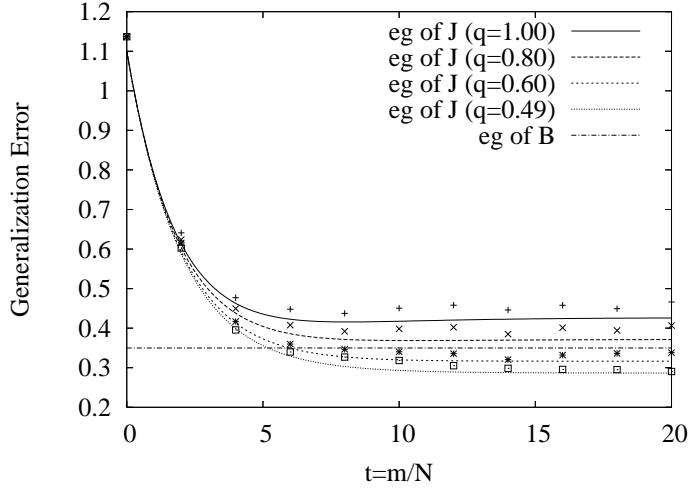}
\caption{Dynamical behaviors of generalization errors 
$\epsilon_{Jg}$. Theory and computer simulations.
Conditions other than $q$ are
$\eta=0.3, K=3, R_B=0.7, \sigma_A^2=0.0, \sigma_B^2=0.1$
and $\sigma_J^2=0.2$.}
\label{fig:egK3RB07E03VA00VB01VJ02}
\end{center}
\end{figure}

\begin{figure}[htbp]
\begin{center}
\includegraphics[width=\gs2\linewidth,keepaspectratio]{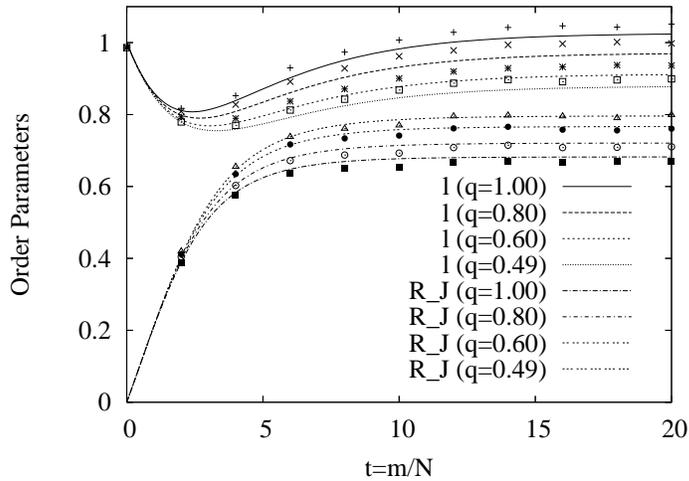}
\caption{Dynamical behaviors of $R_J$ and $l$.
Theory and computer simulations.
Conditions other than $q$ are
$\eta=0.3, K=3, R_B=0.7, \sigma_A^2=0.0, \sigma_B^2=0.1$
and $\sigma_J^2=0.2$.}
\label{fig:RlK3RB07E03VA00VB01VJ02}
\end{center}
\end{figure}

The dynamical behaviors of generalization errors
$\epsilon_{Jg}$
have been analytically obtained by solving 
Eqs. (\ref{eqn:eJg}), (\ref{eqn:defrJ}) and
(\ref{eqn:rBkJ})--(\ref{eqn:barsigma2B2}).
Figure \ref{fig:egK3RB07E03VA00VB01VJ02}
shows the analytical results and the corresponding 
simulation results, where $N=2000$.
In computer simulations, 
$K$ ensemble teachers are used in turn.
$\epsilon_{Jg}$
was obtained by
averaging the squared errors for $10^4$ random inputs 
at each time step.
Generalization error $\epsilon_{Bg}$ of one of 
the ensemble teachers is also shown.
The dynamical behaviors of $R$ and $l$
are shown in Fig. \ref{fig:RlK3RB07E03VA00VB01VJ02}.

In these figures, 
the curves represent theoretical results. 
The dots 
represent simulation results.
Conditions other than $q$
are common:
$\eta=0.3, K=3, R_B=0.7, \sigma_A^2=0.0, \sigma_B^2=0.1$
and $\sigma_J^2=0.2$.
Figure \ref{fig:egK3RB07E03VA00VB01VJ02} shows that
the smaller $q$ is, that is, the richer the variety of the ensemble
teachers is, the smaller 
generalization error $\epsilon_{Jg}$ of
the student is.
Especially in the cases of $q=0.6$ and $q=0.49$,
the generalization error of the student
becomes smaller than a member of 
the ensemble teachers after $t\approx 5$.
This means that
the student in this model 
can become more clever than
each member of the ensemble teachers
even though 
the student only uses the input-output pairs of 
members of the ensemble teachers.
Figure \ref{fig:RlK3RB07E03VA00VB01VJ02} shows that
the larger the variety of the ensemble teachers is,
the larger direction cosine $R_J$ is and
the smaller length $l$ of the student is.
The reason minimum value 0.49 of $q$
is taken as the squared value of $R_B=0.7$
in Figs. \ref{fig:egK3RB07E03VA00VB01VJ02} and
\ref{fig:RlK3RB07E03VA00VB01VJ02}
is described later.

In Figs. \ref{fig:egK3RB07E03VA00VB01VJ02} and
\ref{fig:RlK3RB07E03VA00VB01VJ02}, 
$\epsilon_{Jg}, R_J$ and $l$
almost seem to reach a steady state
by $t=20$.
The macroscopic behaviors of $t \rightarrow \infty$
can be understood theoretically 
since the order parameters have been obtained 
analytically.
Focusing on the signs of the powers of the exponential functions
in Eqs. (\ref{eqn:rBkJ})--(\ref{eqn:l2}),
we can see that
$\epsilon_{Jg}$ and $l$ diverge if $\eta<0$ or $\eta>2$.
The steady state values of 
$r_{BkJ}, r_J$ and $l^2$
in the case of $0<\eta<2$
can be easily obtained by substituting $t\rightarrow \infty$
in Eqs. (\ref{eqn:rBkJ})--(\ref{eqn:l2})
as follows:
\begin{eqnarray}
r_{BkJ} &\rightarrow& q+\frac{1-q}{K}, \label{eqn:strBkJ}\\
r_J &\rightarrow& R_B, \label{eqn:strJ}\\
l^2 &\rightarrow& 
\frac{1}{2-\eta} 
\left( 2\left(1-\eta\right)\left( q+\frac{1-q}{K}\right) 
 +\eta \left(1+\sigma_B^2+\sigma_J^2\right)\right)\\
&=& q+\frac{1-q}{K}+\frac{\eta}{2-\eta}
\left( \frac{(1-q)(K-1)}{K}+\sigma_B^2+\sigma_J^2\right). 
\label{eqn:stl2}
\end{eqnarray}

Equations (\ref{eqn:eJg}), (\ref{eqn:defrJ}) and 
(\ref{eqn:strBkJ})--(\ref{eqn:stl2})
show the following:
in the case of $\eta=1$, the steady value of length $l$
is independent from the number $K$ of teachers
and direction cosine $q$ among the ensemble teachers.
Therefore, the steady value of generalization error
$\epsilon_{Jg}$ and direction cosine $R_J$
are independent from $K$ and $q$ in this case.
In the case of $0<\eta<1$,
the smaller $q$ is or the larger $K$ is,
the smaller the steady values of $l$ and $\epsilon_{Jg}$
are and the larger the steady value of $R_J$ is.
In the case of $1<\eta<2$, on the contrary, 
the smaller $q$ is or the larger $K$ is,
the larger the steady values of $l$ and $\epsilon_{Jg}$
are and the smaller the steady value of $R_J$ is. 
That is,
in the case of $\eta<1$,
the more teachers exist and the richer the variety
of teachers is, the more clever the student can become.
On the contrary, 
in the case of $\eta>1$,
the number of teachers should be small
and 
the variety of teachers should be low
for the student to become clever.

In the right hand side of Eq. (\ref{eqn:stl2}),
since the second and the third terms
are positive,
the steady value of $l$ is larger than $\sqrt{q}$.
In addition, since $l\rightarrow \sqrt{q}$
in the limit of $\eta \rightarrow 0$ and 
$K \rightarrow \infty$,
Eqs. (\ref{eqn:defrJ}) and (\ref{eqn:strJ}) show
$R_J \rightarrow R_B/\sqrt{q}$.
On the other hand,
when $\mbox{\boldmath $S$}$ and 
$\mbox{\boldmath $T$}$ are generated independently
under conditions where the direction cosine 
between $\mbox{\boldmath $S$}$ and $\mbox{\boldmath $P$}$
and 
between $\mbox{\boldmath $T$}$ and $\mbox{\boldmath $P$}$
are both $R_0$, where $\mbox{\boldmath $S$}$, 
$\mbox{\boldmath $T$}$ and $\mbox{\boldmath $P$}$
are high dimensional vectors,
the direction cosine between 
$\mbox{\boldmath $S$}$ and $\mbox{\boldmath $T$}$
is $q_0=R_0^2$, as shown in the appendix.
Therefore, 
if ensemble teachers have enough variety
that they have been generated independently
under the condition that all direction cosines
between ensemble teachers and the true teacher are
$R_B$, $R_B/\sqrt{q}=1$,
then direction cosine $R_J$ between
the student and the true teacher approaches unity
regardless of the variances of noises
in the limit of $\eta \rightarrow 0$ and
$K \rightarrow \infty$.

\begin{figure}[htbp]
\begin{center}
\includegraphics[width=\gs2\linewidth,keepaspectratio]{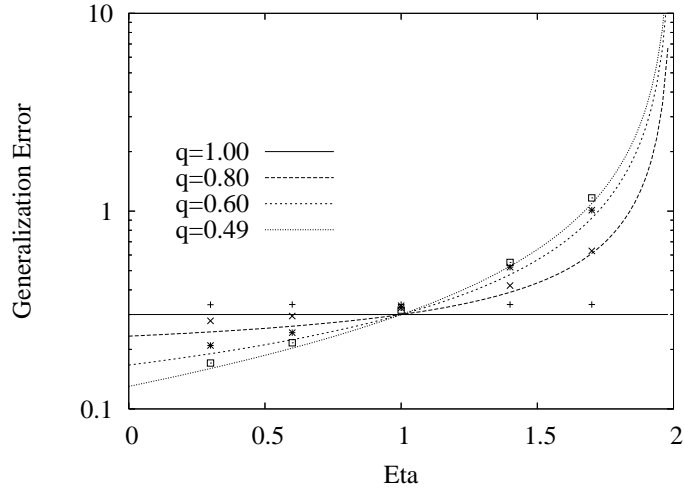}
\caption{Steady value of generalization error $\epsilon_{Jg}$
in the case of $K=3$.
Theory and computer simulations.
Conditions other than $K$ and $q$ are
$\sigma_A^2=\sigma_B^2=\sigma_J^2=0.0$ and $R_B=0.7$.}
\label{fig:egK3RB07VA00VB00VJ00}
\end{center}
\end{figure}

\begin{figure}[htbp]
\begin{center}
\includegraphics[width=\gs2\linewidth,keepaspectratio]{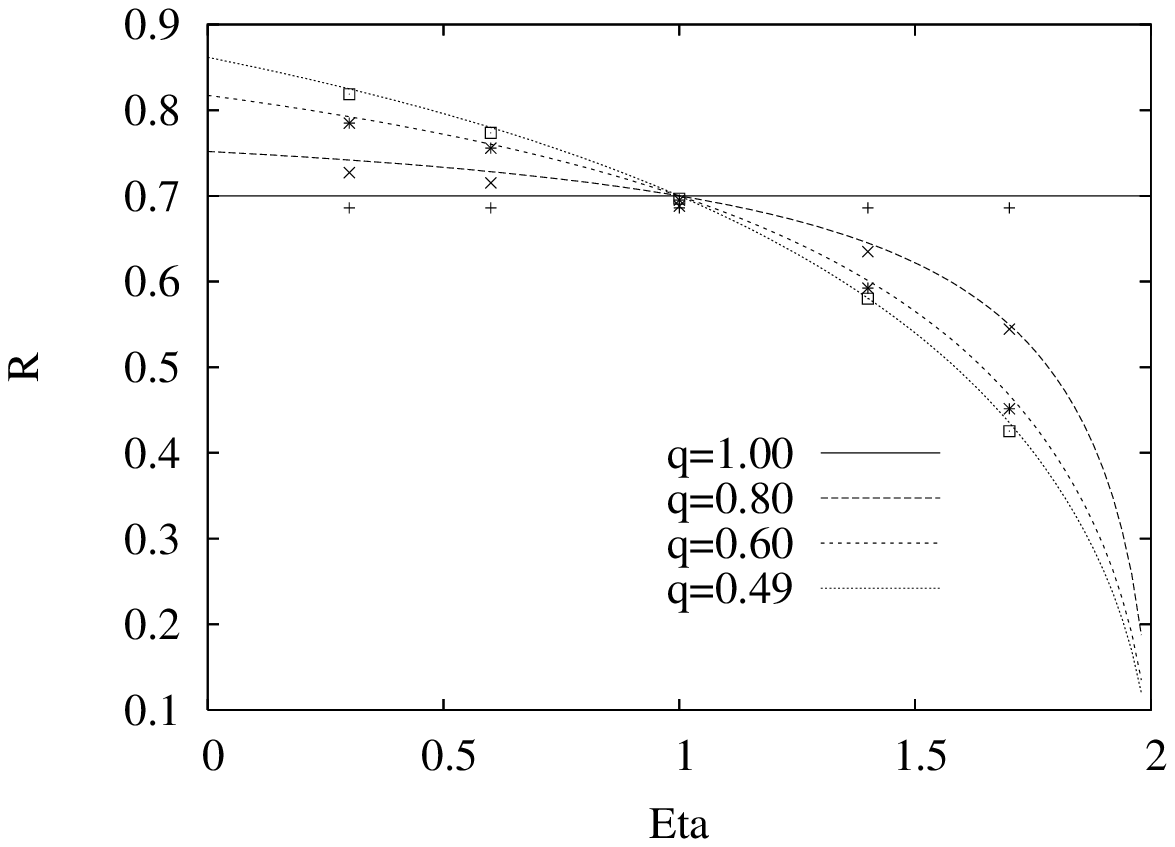}
\caption{Steady value of direction cosine $R_J$
in the case of $K=3$.
Theory and computer simulations.
Conditions other than $K$ and $q$ are
$\sigma_A^2=\sigma_B^2=\sigma_J^2=0.0$ and $R_B=0.7$.}
\label{fig:RK3RB07VA00VB00VJ00}
\end{center}
\end{figure}

\begin{figure}[htbp]
\begin{center}
\includegraphics[width=\gs2\linewidth,keepaspectratio]{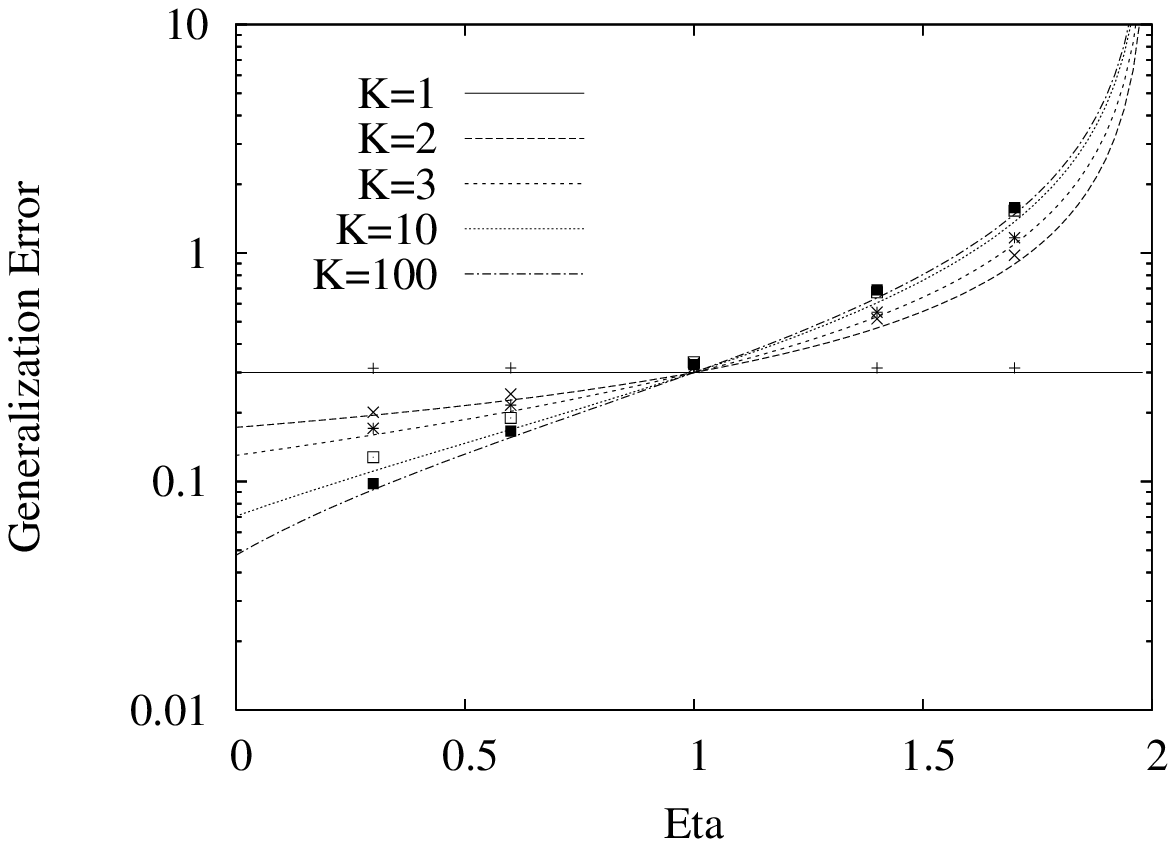}
\caption{Steady value of generalization error $\epsilon_{Jg}$
in the case of $q=0.49$.
Theory and computer simulations.
Conditions other than $K$ and $q$ are
$\sigma_A^2=\sigma_B^2=\sigma_J^2=0.0$ and $R_B=0.7$.}
\label{fig:egq049RB07VA00VB00VJ00}
\end{center}
\end{figure}

\begin{figure}[htbp]
\begin{center}
\includegraphics[width=\gs2\linewidth,keepaspectratio]{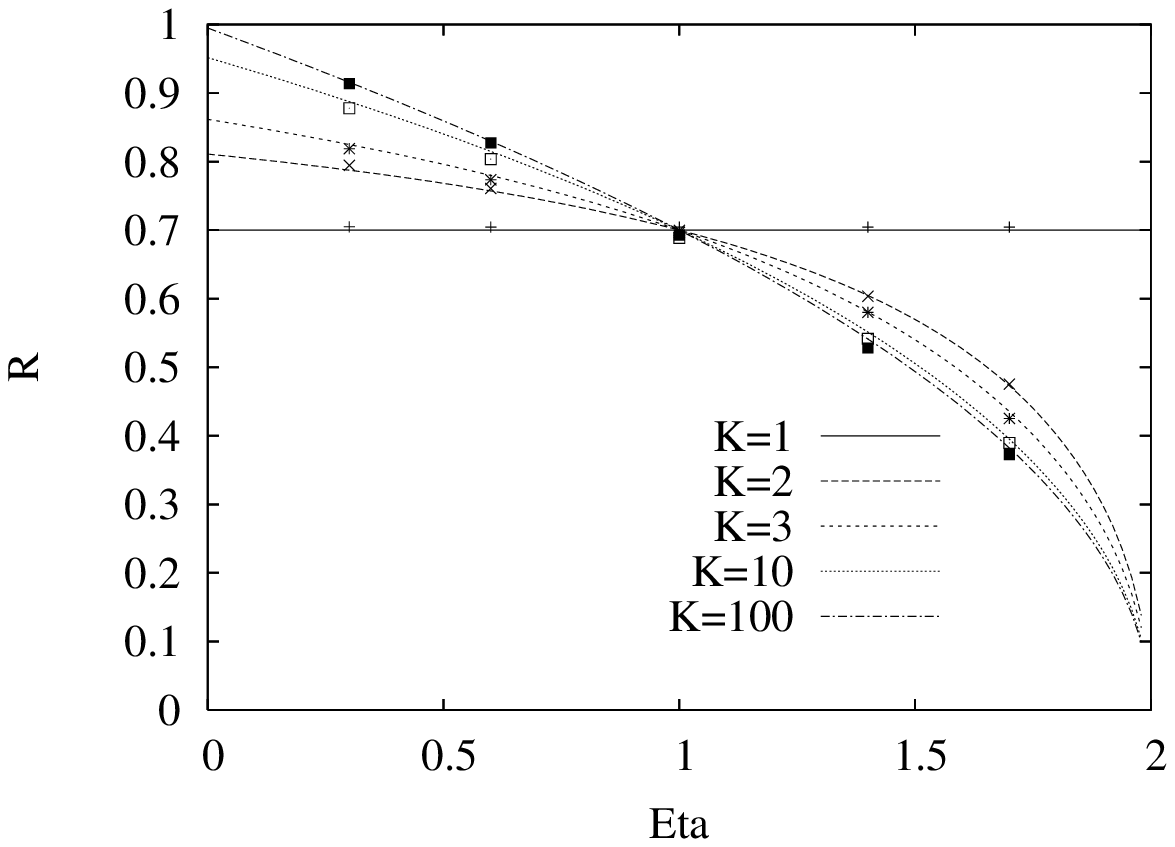}
\caption{Steady value of direction cosine $R_J$
in the case of $q=0.49$.
Theory and computer simulations.
Conditions other than $K$ and $q$ are
$\sigma_A^2=\sigma_B^2=\sigma_J^2=0.0$ and $R_B=0.7$.}
\label{fig:Rq049RB07VA00VB00VJ00}
\end{center}
\end{figure}

Figures \ref{fig:egK3RB07VA00VB00VJ00}--\ref{fig:Rq049RB07VA00VB00VJ00}
show 
the relationships between learning rate $\eta$
and $\epsilon_{Jg}$, $R_J$.
In Figs \ref{fig:egK3RB07VA00VB00VJ00}
and \ref{fig:RK3RB07VA00VB00VJ00},
$K=3$ and is fixed.
In Figs \ref{fig:egq049RB07VA00VB00VJ00}
and \ref{fig:Rq049RB07VA00VB00VJ00},
$q=0.49$ and is fixed.
Conditions other than $K$ and $q$
are 
$\sigma_A^2=\sigma_B^2=\sigma_J^2=0.0$ and $R_B=0.7$.
Computer simulations have been executed using
$\eta=0.3, 0.6, 1.0, 1.4$ and $1.7$.
The values on $t=20$ are plotted for the simulations 
and considered to have already reached a steady state.

These figures show the following:
the smaller learning rate $\eta$ is,
the smaller generalization error $\epsilon_{Jg}$ is
and the larger direction cosine
$R_J$ is.
Needless to say,
when $\eta$ is small, learning is slow.
Therefore, residual generalization error 
and learning speed are in a relationship tradeoff.
The phase transition in which
$\epsilon_{Jg}$ diverges 
and $R_J$ becomes zero on $\eta=2$
is shown.
In the case of $\eta<1$,
the larger $K$ is
or the smaller $q$ is, that is, the richer the variety of 
ensemble teachers is, 
the smaller $\epsilon_{Jg}$ is 
and the larger $R_J$ is.
On the contrary, 
the properties are completely reversed
in the case of $\eta>1$.

\begin{figure}[htbp]
\begin{center}
\includegraphics[width=\g3\linewidth,keepaspectratio]{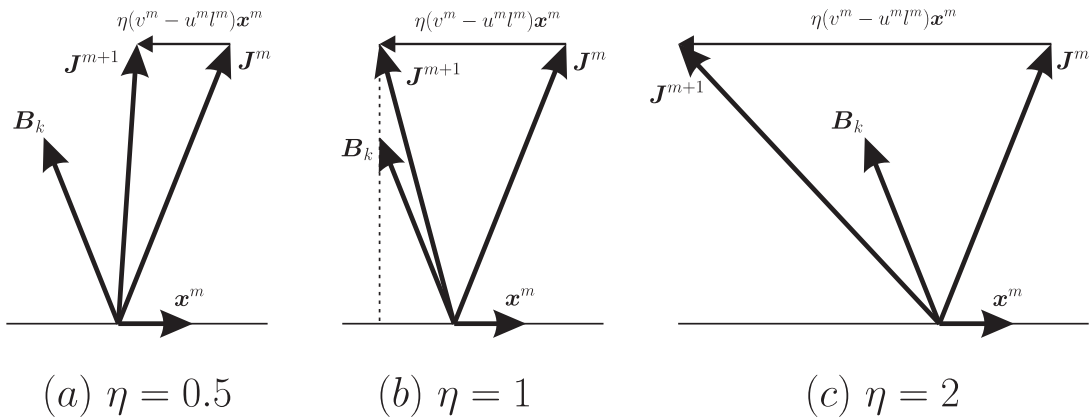}
\caption{Geometric meaning of learning rate $\eta$}
\label{fig:update}
\end{center}
\end{figure}

As described above,
learning properties are dramatically changed with 
learning rate $\eta$.
It is difficult to explain the reason qualitatively.
Here, we try to explain the reason intuitively 
by showing the geometrical meaning of $\eta$.
Figures \ref{fig:update}(a)--(c)
show the updates of $\eta=0.5$, $\eta=1$ and $\eta=2$,
respectively.
Here, the noises are ignored for simplicity.
Needless to say, 
teacher $\mbox{\boldmath $B$}_k$ itself
cannot be observed directly and
only output $v$ can be observed
when student $\mbox{\boldmath $J$}$ is updated.
In addition, 
since the projections from $\mbox{\boldmath $J$}^{m+1}$
to $\mbox{\boldmath $x$}^m$
and from $\mbox{\boldmath $B$}_k$
to $\mbox{\boldmath $x$}^m$ are equal
in the case of $\eta=1$, as shown in 
Fig. \ref{fig:update}(b),
$\eta=1$
is a special condition
where the student uses up the information
obtained from input $\mbox{\boldmath $x$}^m$.
In the case of $\eta<1$,
the update is short.
Since in a sense this fact helps balance the information
from the ensemble teachers,
the generalization error of the student 
is improved when 
the number $K$ of teachers is large and
their variety is rich.
On the other hand, 
the update is excessive
when $\eta>1$.
Therefore, the student is shaken or swung,
and its generalization performance worsens
when $K$ is large and the variety is rich.
In addition, 
the reason that learning diverges 
if $\eta<0$ or $\eta>2$
can be understood intuitively
from Fig. \ref{fig:update}:
distance 
$\|(\eta-1)(v^m-u^ml^m)\mbox{\boldmath $x$}^m\|$,
measured by 
the projections to $\mbox{\boldmath $x$}^m$
between student $\mbox{\boldmath $J$}^{m+1}$
after the update and teacher $\mbox{\boldmath $B$}_k$,
is larger than 
distance $\|(v^m-u^ml^m)\mbox{\boldmath $x$}^m\|$
between student $\mbox{\boldmath $J$}^m$
before the update and teacher $\mbox{\boldmath $B$}_k$
in the case of $\eta<0$ or $\eta>2$.
Therefore, the learning diverges.

\section{Conclusion}
We analyzed the generalization performance of a student 
in a model composed of linear perceptrons: 
a true teacher, ensemble teachers, and the student. 
The generalization error 
of the student was analytically calculated 
using statistical mechanics
in the framework of online learning, 
proving that
when learning rate $\eta <1$,
the larger the number $K$ and the variety of 
the ensemble teachers are,
the smaller the generalization error is.
On the other hand, when $\eta >1$,
the properties are completely reversed.
If the variety of ensemble teachers is rich enough, 
the direction cosine between the true teacher and the 
student becomes unity in the limit of $\eta \rightarrow 0$ and
$K \rightarrow \infty$.

\section*{Acknowledgments}
This research was partially supported by the Ministry of Education, 
Culture, Sports, Science, and Technology of Japan, 
with Grants-in-Aid for Scientific Research
14084212, 15500151 and 16500093.

\appendix
\section{Direction cosine $q$ among ensemble teachers}
\begin{figure}[htbp]
\vspace{3mm}
\begin{center}
\includegraphics[width=\gsize\linewidth,keepaspectratio]{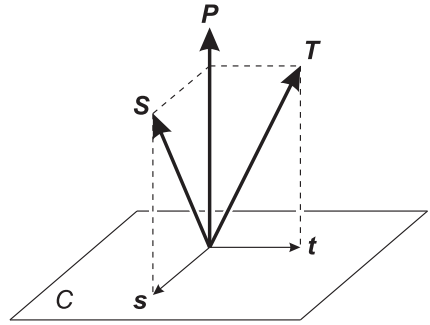}
\caption{Direction cosine among ensemble teachers}
\label{fig:PST}
\end{center}
\end{figure}

Let us consider the case where
$\mbox{\boldmath $S$}$ and 
$\mbox{\boldmath $T$}$ are generated independently
satisfying the condition that direction cosines 
between $\mbox{\boldmath $S$}$ and $\mbox{\boldmath $P$}$
and 
between $\mbox{\boldmath $T$}$ and $\mbox{\boldmath $P$}$
are both $R_0$,
as shown in Fig. \ref{fig:PST}, 
where $\mbox{\boldmath $S$}$, 
$\mbox{\boldmath $T$}$ and $\mbox{\boldmath $P$}$
are $N$ dimensional vectors.
In this figure, 
the inner product of 
$\mbox{\boldmath $s$}$ and
$\mbox{\boldmath $t$}$ is
\begin{eqnarray}
\mbox{\boldmath $s$}\cdot\mbox{\boldmath $t$}
&=& 
\left(\mbox{\boldmath $S$}-
R_0
\frac{\|\mbox{\boldmath $S$}\|}{\|\mbox{\boldmath $P$}\|}
\mbox{\boldmath $P$}\right)
\cdot
\left(\mbox{\boldmath $T$}-
R_0
\frac{\|\mbox{\boldmath $T$}\|}{\|\mbox{\boldmath $P$}\|}
\mbox{\boldmath $P$}\right)\\
&=&
\|\mbox{\boldmath $S$}\|\|\mbox{\boldmath $T$}\|
\left(q_0-R_0^2\right),
\end{eqnarray}
where $\mbox{\boldmath $s$}$ and $\mbox{\boldmath $t$}$ 
are
projections from 
$\mbox{\boldmath $S$}$ to
the orthogonal complement $C$ of $\mbox{\boldmath $X$}$
and 
from 
$\mbox{\boldmath $T$}$ to $C$, respectively.
$q_0$ denotes the direction cosine between
$\mbox{\boldmath $S$}$ and $\mbox{\boldmath $T$}$.

Incidentally, 
if dimension $N$ is large and 
$\mbox{\boldmath $S$}$ and $\mbox{\boldmath $T$}$
have been generated independently,
$\mbox{\boldmath $s$}$ and $\mbox{\boldmath $t$}$
should be orthogonal to each other.
Therefore, $q_0=R_0^2$.

\end{document}